\documentclass[twocolumn]{aastex701}

\usepackage{amsmath}
\usepackage{hyperref}

\usepackage{xcolor}

\begin{document}

\title{The CGM with local universe FRBs: evidence of strong AGN feedback in a massive elliptical galaxy}

\author[orcid=0009-0008-5043-6220]{Samuel McCarty}
\affiliation{Center for Astrophysics $|$ Harvard \& Smithsonian}
\email[show]{samuel.mccarty@cfa.harvard.edu}  

\author[orcid=0000-0002-7587-6352]{Liam Connor} 
\affiliation{Center for Astrophysics $|$ Harvard \& Smithsonian}
\email{liam.connor@cfa.harvard.edu}

\author[orcid=0000-0001-8235-2939]{Ralf M. Konietzka}
\affiliation{Center for Astrophysics $|$ Harvard \& Smithsonian}
\email{ralf.konietzka@cfa.harvard.edu}

\begin{abstract}

Modern cosmology and galaxy formation rely on an understanding of how cosmic baryons are distributed, a significant portion of which exist in the diffuse gas confined to halos. Fast Radio Bursts (FRBs) are a promising probe of the Universe's ionized gas. At low redshift, the contribution to the dispersion measure (DM) from the intergalactic medium (IGM) and intervening halos is subdominant, allowing us to study the circumgalactic media (CGM) of the host galaxies, if the host interstellar medium (ISM) can be sufficiently suppressed. We select a sample of five local universe FRBs with host halos spanning a wide range in mass ($\sim10^{11-13}\,M_\odot$), and with indicators that host ISM DM is negligible. We use these FRBs to constrain the CGM in each halo by comparing to simulations and measuring the CGM mass assuming a range of spherical gas profiles. We find one of our sources, the only massive elliptical host galaxy, strongly disfavors low feedback scenarios in simulations, and has been evacuated of its baryons ($\sim$\,10$\%$ of the cosmological average $\frac{\Omega_b}{\Omega_m}$ within $R_\mathrm{vir}$). This galaxy shows evidence of a past episode of AGN activity, consistent with the picture of strong AGN feedback in galaxy group-scale halos. The gas mass measurements for the other sources tentatively support more baryon retention in $L_*$ galaxies compared to group-scale halos. Finally, we demonstrate that a large sample of local universe FRBs will enable precision measurements of halo gas.

\end{abstract}

\section{Introduction}

The nature of the gas gravitationally bound to dark matter halos is a key limiting factor in our understanding of galaxy formation and cosmology. For galaxies, this circumgalactic medium (CGM) acts as an interface with the galaxy's cosmological environment. It plays a key role in the baryon cycle, which controls the growth of the galaxy through accretion from the intergalactic medium (IGM), feedback processes, and recycling \citep{Tumlinson_2017}. 
For cosmology, the suppression of the matter power spectrum due to baryonic effects on small scales is a leading source of uncertainty in the upcoming Stage IV weak lensing surveys by the \textit{Euclid} satellite \citep{EuclidI}, 
the Vera C. Rubin Observatory \citep{Ivezi_2019},
and the \textit{Nancy Roman Space Telescope} \citep{spergel2013widefieldinfraredsurveytelescopeastrophysics}, 
thus complicating our understanding of dark matter and dark energy (e.g. \cite{Semboloni_2011}). 
Further, the relative amount of baryons in different components of the diffuse gas (the CGM, IGM, intragroup medium, intracluster medium) has been debated for decades -- often referred to as  ``missing baryons problem'' \citep{Cen1999}. 

While traditional probes of halo gas have made significant progress this decade, they have several drawbacks. It is difficult to detect low to intermediate-mass halos in X-ray or the Sunyaev-Zeldovich (SZ) effect \citep{SZoriginal}. The thermal Sunyaev-Zeldovich effect (tSZ) and X-rays further rely on assumptions about gas temperature and metallicity, respectively. The kinetic Sunyaev-Zeldovich (kSZ) signal is intrinsically faint, such that it has only been measured for galaxy group-scale and larger halos (e.g. \cite{Schaan_2021}, \cite{Guachalla_2025}). Quasar absorption spectroscopy relies on assumptions about ionization state and structure of the gas \citep{Tumlinson_2017}.

On the other hand, Fast Radio Bursts (FRBs) have emerged as a promising new method of studying the gas in and around halos. FRBs are millisecond radio transients whose signals are dispersed as they travel through cosmic plasma \citep{Lorimer_2007, Petroff_2019, Cordes_2019}. The dispersion measure (DM) of FRBs is a clean probe of electron column density along the line of sight \citep{Petroff_2019}:

\begin{equation}
    \mathrm{DM}(z)=\int_0^z \frac{n_e(z')c}{(1+z')^2H(z')}dz',
    \label{eqn:DM_z}
\end{equation}

\noindent where $n_e$ is the electron density and  $H(z)$ is the Hubble parameter. 

Significant progress has been made towards using FRBs as astrophysical and cosmological tools. One approach studies the distribution of observed DM as a function of redshift to constrain the IGM and feedback  \citep{McQuinn_2013,Macquart_2020,reischke2025measurementbaryonicfeedbackfast,Connor2025,Sharma_2025}. Others correlate the FRB signal with various observables \citep{reischke2024calibratingbaryonicfeedbackweak,Hussaini_2025,wang2025measurementdispersionunicodex2013galaxycrosspowerspectrum,leung2025nullingbaryonicfeedbackweak,sharma2025probingbaryonicfeedbackcosmology,takahashi2025measurementangularcrosscorrelationcosmological,leung2025stellarmassdispersionmeasurecorrelations,Hsu2025}. However, much work remains to be done towards studying the gas in halos with FRBs. One promising approach is to use FRBs intersecting intervening halos along the line of sight \citep{Connor2021, Wu_2023,Khrykin_2024,Hussaini_2025}, although this is limited by the current FRB sample size.
 There have been a few investigations of groups/clusters \citep{Connor_2023,leeGC,lanman2025constrainingbaryonfractionsgalaxy} and targeted studies of FRBs intersecting M31 \citep{ProchaskaZheng2019, connor_2020, Anna-Thomas_2025,kahinga2026constrainingmassm31ionized}. \cite{leung2025stellarmassdispersionmeasurecorrelations} investigated the relationship between host galaxy DM and stellar mass for a sample of local universe FRBs. They found an anticorrelation, which aligns more closely with simulations that have stronger feedback implementations, constraining feedback in $M_\mathrm{h}\sim10^{11-13}\,M_\odot$ halos.

The Milky Way (MW) CGM itself is an obvious target, and recent works have placed upper limits on its integrated DM  \citep{Kirsten_2022,Cook_2023,Ravi_2025}. However, these studies were dominated by individual low-DM  FRBs. Utilizing larger samples of FRBs, such as the recent work by \cite{hoffmann2026ihaloconstrainingmilky}, will enable a more comprehensive characterization of the MW CGM, and may enable detection of anisotropies \citep{ProchaskaZheng2019, liu2026investigatinganisotropydispersionmeasure}. 

In addition to the MW, FRBs offer a critical opportunity to study the CGM in $10^{11-13}M_\odot$ halos which has so far been largely untapped. In this work, we propose to use local universe FRBs for this purpose. Below $z\lesssim0.2$, the DM budget is no longer dominated by the cosmic web, allowing the contribution from the CGM of the host to be more easily isolated from that of the cosmic web, leaving the host ISM as the main remaining uncertainty.

Because it is not in general possible to disentangle the DM from the host interstellar medium (ISM) and host CGM, we select a sample of FRBs for which there is reason to believe that ISM DM is negligible 
(Section \ref{sec:sample}). We analyze this sample in Section \ref{sec:analysis} to determine the gas content of the host halos.  In Section \ref{sec:discussion} we provide a discussion of our results and future prospects for this method in light of upcoming all-sky monitors. Throughout this work we assume the best-fit Planck 2018+BAO cosmology \citep{Plank18}.

\section{FRB sample and DM accounting}
\label{sec:sample}

\begin{figure*}
    \centering
    \includegraphics[width=\linewidth]{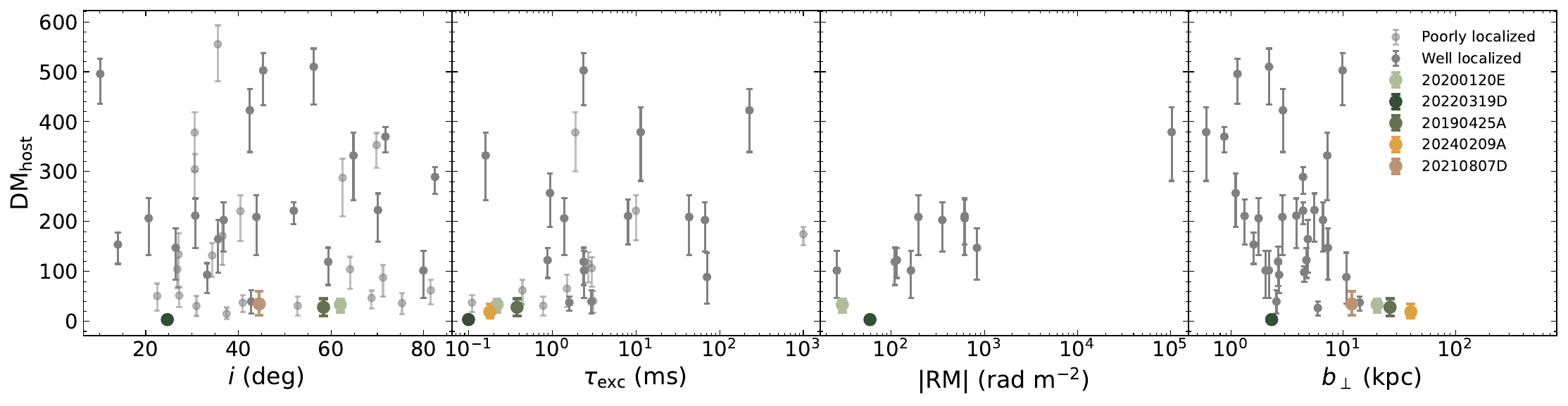}
    \caption{$\mathrm{DM_{host}}$ for all localized FRBs with $z<0.2$ at the time of writing plotted against potential indicators of $\mathrm{DM_{host,ISM}}$: host galaxy inclination $i$, scattering timescale $\tau_\mathrm{exc}$, rotation measure RM, and impact parameter $b_\perp$. The colored points are the five FRBs selected for our analysis. Light grey points are the poorly localized ($\sim$arcmin) FRBs. }
    \label{fig:sample}
\end{figure*}

\begin{figure*}
    \centering
    \includegraphics[width=\linewidth]{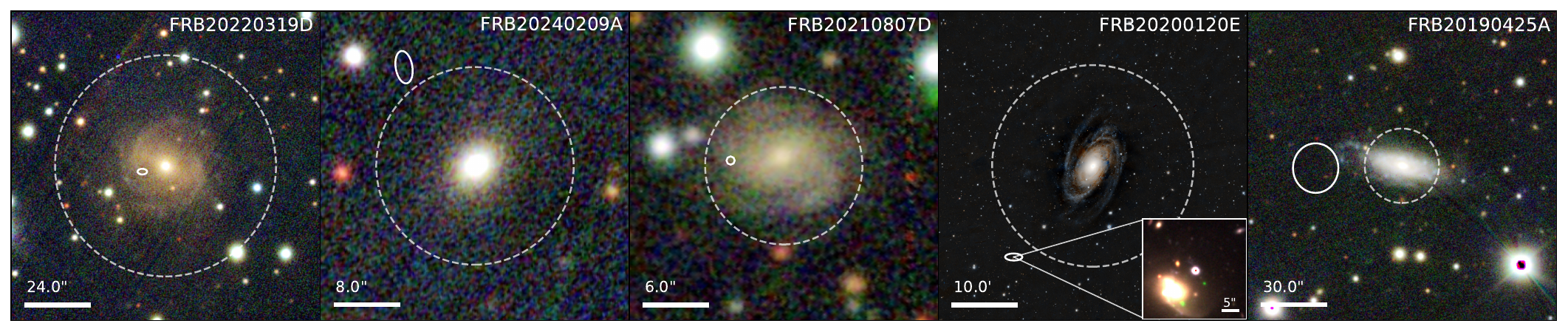}
    \caption{Pan-STARRS1 \citep{chambers2019panstarrs1surveys} gri color images of the FRB host galaxies in our sample. The FRB localization regions are shown as solid white ellipses. The dashed white circles are 0.1$R_\mathrm{500c}$ for each galaxy's halo. For FRB20200120E, the inset shows the VLBI localization to a globular cluster in M81 presented by \cite{Kirsten_2022}.}
    \label{fig:images}
\end{figure*}

\begin{figure*}
    \centering
    \includegraphics[width=\linewidth]{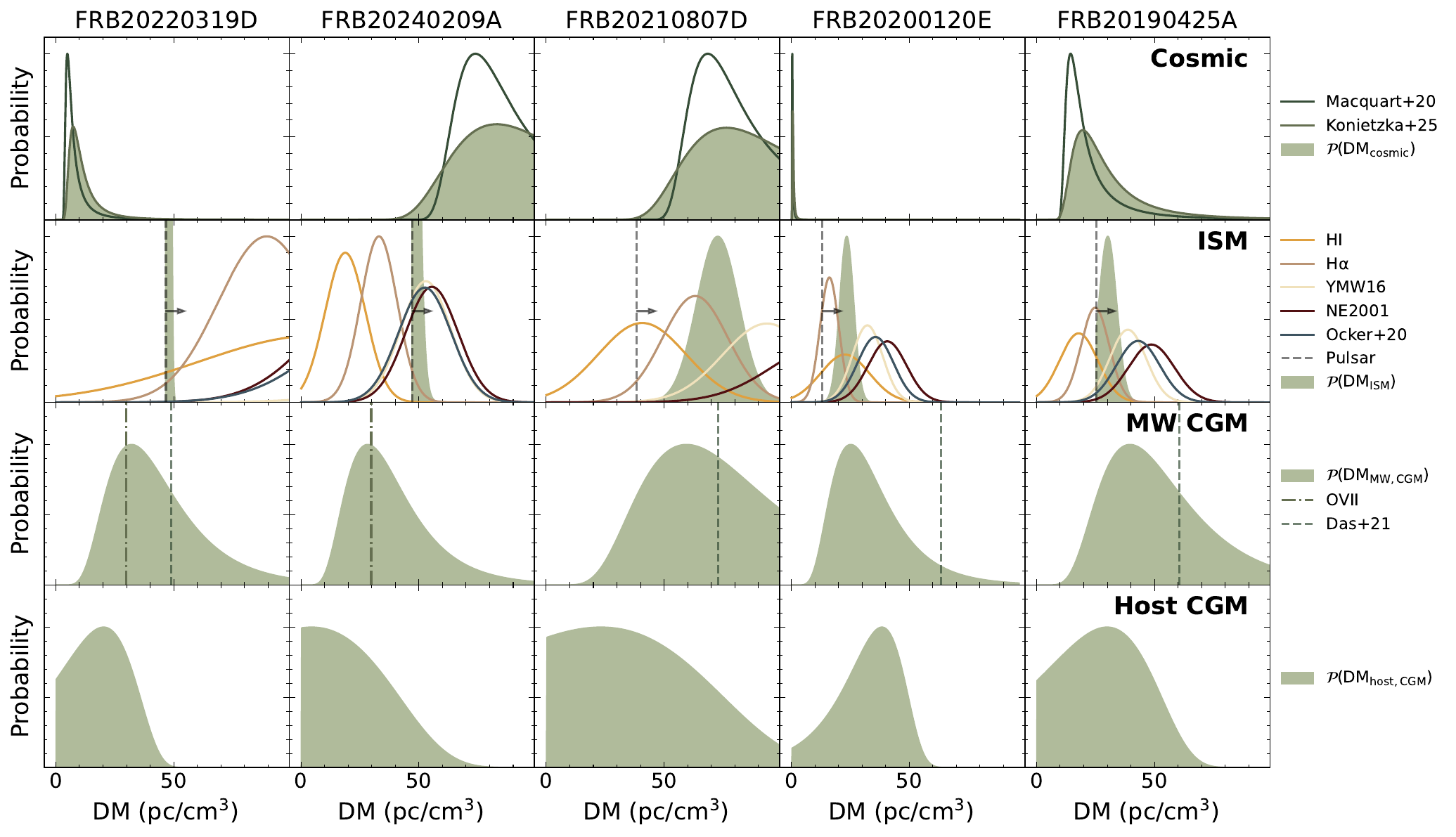}
    \caption{The PDFs of the different DM components using various estimates, with the shaded curves showing the adopted PDFs used in our analysis. In each plot, all curves are arbitrarily scaled by the same factor for better visualization. The top row shows $p(\mathrm{DM_{cosmic}})$, which is estimated using the best fit of \cite{konietzka2025raytracingfastradiobursts}, with the relation of \cite{Macquart_2020} plotted for reference. The middle row shows $p(\mathrm{DM_{MW,ISM}})$, where the final PDF is the combination of the HI, H$\alpha$, and YMW16 \citep{Yao_2017} estimates, truncated by the pulsar lower limit (vertical dashed line). For FRB 20220319D, the mean $\mathrm{DM_{MW,ISM}}$ estimate is much larger than $\mathrm{DM_{obs}}$, so we set $p(\mathrm{DM_{MW,ISM}})$ as a delta function at the pulsar lower limit following \citep{Ravi_2025}. We also show the NE2001 \citep{cordes2003ne2001inewmodelgalactic,Ocker_2024} and \cite{Ocker_2020} models for comparison. The third row shows  YT20 \citep{Yamasaki_2020} estimate of $\mathrm{DM_{MW,CGM}}$. We also plot an estimate of $\mathrm{DM_{MW,CGM}}$ using the nearest OVII absorption sightline to each FRB from \cite{Fang_2015} and the scaling relation between $N_\mathrm{OVII}$ and DM from \citep{Yamasaki_2020}, as well as the median of sightlines within 20$^{\circ}$ from \cite{Das2021} (but note that the uncertainty in both of these is quite large). In the bottom panel, we plot the final $p(\mathrm{DM_{host}})$, which under our assumption of negligible $\mathrm{DM_{host,ISM}}$ becomes $p(\mathrm{DM_{host,CGM}})$.}
    \label{fig:pdfs}
\end{figure*}

The observed DM of an FRB can be divided into several parts:

\begin{equation}
    \begin{split}
        \mathrm{DM_{obs}}=\mathrm{DM_{MW,ISM}}+\mathrm{DM_{MW,CGM}}+\mathrm{DM_{cosmic}} + \\ +\frac{\mathrm{DM_{host}}}{1+z}
    \end{split}
    \label{eqn:DMobs}
\end{equation}

\begin{equation}
    \mathrm{DM_{host}}=\mathrm{DM_{host,ISM}}+\mathrm{DM_{host,CGM}}.
\end{equation}

\noindent $\mathrm{DM_{MW,ISM}}$ and $\mathrm{DM_{MW,CGM}}$ are the contributions from the gas in the MW's ISM and CGM, respectively, and $\mathrm{DM_{host,x}}$ refers to the same for the FRB host galaxy. $\mathrm{DM_{cosmic}}$ is the contribution from the IGM and intervening halos along the line of sight. Note that some authors distinguish further a DM component from circumburst material in the local environment of the FRB, we group this into $\mathrm{DM_{host,ISM}}$. 

We wish to isolate the $\mathrm{DM_{host,CGM}}$ term. Under the assumption that different components of the DM are independent, which is not in general true but a good approximation (see e.g. \citep{konietzka2025raytracingfastradiobursts} for the correlation between $\mathrm{DM_{host}}$ and $\mathrm{DM_{cosmic}}$), the probability density function (PDF) can be written:
\begin{equation}
    \begin{split}
        p\left(\mathrm{\frac{DM_{host,CGM}}{1+z}}\right)=p(\mathrm{DM_{obs}})\ast p(\mathrm{-DM_{MW,ISM}}) \\ \ast p(\mathrm{-DM_{MW,CGM}}) \ast p(\mathrm{-DM_{cosmic}}) \ast p\left(\mathrm{\frac{-DM_{host,ISM}}{1+z}}\right).
    \end{split}
\end{equation}

\subsection{Cosmic DM}

In previous studies the contribution of the cosmic web towards the total DM, $p(\mathrm{DM_{cosmic}}|z)$, has been modeled using the parameterization presented in \cite{Macquart_2020}.
However, recent work suggests that this prescription of $p(\mathrm{DM_{cosmic}}|z)$ does not fully capture the higher-order moments of the distribution \citep{konietzka2025raytracingfastradiobursts, Sharma_2025}.
Therefore, in this work, we use the best-fit functional form from \cite{konietzka2025raytracingfastradiobursts} for $p(\mathrm{DM_{cosmic}}|z)$, which is described by a modified-normal density function with a power-law tail at the high DM end:
\begin{equation}\label{eq:pdf_cosmic}
    p(\mathrm{DM_{cosmic}}|z) = c\exp\left(-\frac{\mu^2\left(\left(\frac{\mu}{x}\right)^\alpha - 1\right)^2}{2\,\sigma^2\,\alpha^2}\right)\left(\frac{\mu}{x}\right)^{\beta} 
\end{equation}
where the normalization constant $c$ can be obtained via
\begin{equation}\label{eq:pdf_cosmic_norm}
    c^{-1} = \frac{\mu}{\alpha}\,e^{-\frac{\eta^2}{2}}\,\left(\frac{\alpha\,\sigma}{\mu}\right)^\delta\,\Gamma\left(\delta\right)\,\mathrm{D}_{-\delta}\left(-\sqrt{2}\,\eta\right),
\end{equation}
where $\delta = \frac{\beta-1}{\alpha}$, $\eta^2 = \frac{\mu^2}{2\,\sigma^2\,\alpha^2}$, $\Gamma\left(\cdot\right)$ is the gamma function and $D_\cdot(\cdot)$ the parabolic cylinder function.
As in \cite{konietzka2025raytracingfastradiobursts}, we use shape parameters $\alpha \approx 1$ and $\beta\approx3.3$.
The remaining two parameters $\mu$ and $\sigma$ are tuned using simulated DM \citep{konietzka2025raytracingfastradiobursts} drawn from the IllustrisTNG simulation \citep{nelson2021illustristngsimulationspublicdata}.
To this end, we make use of the continuous DM catalog\footnote{The continuous DM catalog along with other DM maps and catalogs from \cite{konietzka2025raytracingfastradiobursts} can be downloaded at \href{https://ralfkonietzka.github.io/fast-radio-bursts/ray-tracing-catalogs/}{https://ralfkonietzka.github.io/fast-radio-bursts/ray-tracing-catalogs/}.}   provided in \cite{konietzka2025raytracingfastradiobursts} that traces rays through the IllustrisTNG simulation while reconstructing all traversed line segments within the underlying Voronoi mesh.
Although this relation is fit to the IllustrisTNG simulation, we do not expect this to bias our results at the low redshifts considered in this work, as the probability of intersecting a foreground halo as well as the total expected IGM contribution for our local Universe sources is low. Similarly, the model of \cite{Sharma_2025} will not differ significantly from our assumed model at these low redshifts.

\subsection{Milky Way DM}
At low redshift it is important to consider the impact of our choice of model for $p(\mathrm{DM_{MW,ISM}})$, which could significantly bias our results. Popular choices for $p(\mathrm{DM_{MW,ISM}})$ are the NE2001 \citep{cordes2003ne2001inewmodelgalactic, Ocker_2024} and YMW16 \cite{Yao_2017} models, which were fit to pulsar DM data. However, as seen in \cite{Ravi_2025}, these models can overestimate $\mathrm{DM_{MW,ISM}}$, especially at low galactic latitudes. Because we expect $\mathrm{DM_{host,CGM}}$ to be $\mathcal{O}(10-100)$ pc/cm$^3$, we need to be more conservative. For this reason we follow \cite{Ravi_2025} and consider independent estimates of $\mathrm{DM_{MW,ISM}}$. First, we estimate $\mathrm{DM_{MW,ISM}}$ for a given FRB sightline using the scaling between the neutral hydrogen column $N_\mathrm{H}$ and DM found by \cite{He_2013}. The column densities are retrieved from the HI4PI survey \citep{HI4PI}. The associated uncertainty comes from the scatter in the relationship. Next, we use the H$\alpha$ emission measure (EM) to estimate $\mathrm{DM_{MW,ISM}}$. The H$\alpha$ flux data is obtained from the Wisconsin H$\alpha$ Mapper (WHAM) \citep{Haffner_2003}, and the analysis follows \cite{Berkhuijsen_2006}. The H$\alpha$ flux is corrected for dust reddening and converted to an EM with associated uncertainty assuming a mean electron temperature $T_e=8000\,$K. Because $\mathrm{DM}\propto n_e l$, where $l$ is the distance, and $\mathrm{EM}\propto n_e^2 l$, we can estimate $\mathrm{DM}\sim \sqrt{\mathrm{EM}l}$, with some assumption about the structure of the gas for which we follow \cite{Ravi_2025}. To estimate the distance through the ISM for extragalactic FRBs we assume the ISM is a disk with height $1.25\,$kpc (\cite{Berkhuijsen_2006} find the average electron density $\langle n_e\rangle$ is constant to a height of $\sim1\,$kpc and then drops off with a sale length $\sim0.3\,$kpc), and an uncertainty on this distance of 20\%. 

With the HI and H$\alpha$ estimates in hand, we combine these with the pulsar-based estimates, the NE2001 and YMW16 models (assuming 20\% uncertainty on each), to get a final distribution $p(\mathrm{DM_{MW,ISM}})$. We further identify the closest pulsar to each FRB sightline within 1 degree of galactic latitude from the ANTF pulsar catalogue\footnote{\url{https://www.atnf.csiro.au/research/pulsar/psrcat}} \citep{Manchester_2005}, excluding globular cluster pulsars, and use the DM of the pulsar as a hard lower limit by truncating $p(\mathrm{DM_{MW,ISM}})$. Each FRB is within 10 degrees of the nearest pulsar. In a special case for FRB 20220319D, the $\mathrm{DM_{MW,ISM}}$ estimate is much larger than the total $\mathrm{DM_{obs}}$, so we follow \cite{Ravi_2025} and use only the pulsar lower limit instead (a delta function for $p(\mathrm{DM_{MW,ISM}})$). By combining multiple independent estimates, we aim to mitigate the risk of systematically overestimating  $\mathrm{DM_{MW,ISM}}$. 

To estimate $\mathrm{DM_{MW,CGM}}$ we use the model of \cite{Yamasaki_2020} (hereafter YT20), which is fit to X-ray EM data. In general there is a lack of independent data available to constrain this component, we take the YT20 model as a conservative estimate, and discuss the impact of $\mathrm{DM_{MW,CGM}}$ in Section \ref{sec:mw}. We include a lognormal scatter of 0.2\,dex around the YT20 value for each FRB sightline.

\subsection{Host DM}

It remains to separate the $\mathrm{DM_{host,ISM}}$ and $\mathrm{DM_{host,CGM}}$ components. Our approach is as follows: we identify FRB and host galaxy properties that are well known to be indicators of $\mathrm{DM_{host,ISM}}$, and use these to select a sample of FRBs for which there is reason to believe $\mathrm{DM_{host,ISM}}$ is small. Under the ansatz that $\mathrm{DM_{host,ISM}}$ is negligible, we obtain results that are consistent with existing data. In the future, it may be possible to estimate $\mathrm{DM_{host,ISM}}$ independently (from resolved integral-field unit observations of H$\alpha$ in the host galaxy for example), but it is not feasible with available data. While unlikely, it is possible that $\mathrm{DM_{host,ISM}}$ is non-negligible for a given source in our sample. However, this would serve only to decrease the budget for $\mathrm{DM_{host,CGM}}$, tightening constraints and strengthening the conclusions we draw from our analysis. See Section \ref{sec:hostism} for a discussion about the impact of host ISM on our analysis.

Several properties of the FRBs and their host galaxies are indicative of low $\mathrm{DM_{host,ISM}}$. In particular, we identify the host galaxy inclination $i$, the FRB scattering timescale $\tau_\mathrm{exc}$ (with MW contribution subtracted), the FRB rotation measure (RM), and the impact parameter of the FRB to the host galaxy $b_\perp$ as promising indicators. An FRB originating from the disk of a face-on spiral galaxy will have traveled less distance through the host ISM than an edge-on spiral, so we expect $i$ to correlate with $\mathrm{DM_{host,ISM}}$ \citep{bhardwaj2024selectionbiasobfuscatesdiscovery}. Similarly, we expect the local environment and host ISM to be the main contributors to the scattering and RM of the FRB. This is because the ISM and circumstellar material can produce density fluctuations appropriate for Faraday rotation and strong scattering; in the IGM \citep{Luan_2014} and CGM this is much less likely \citep{vedanthamphinney}. Finally, the farther away from the host galaxy the FRB originates, the less ISM gas it will encounter. Indeed, \cite{Li_2025} find a correlation between extragalactic DM and $\mathrm{cos}i$, $\tau_\mathrm{exc}$, and $b_\perp$. A similar perscription for selection cuts on $i$ and $\tau_\mathrm{exc}$ was used in \cite{leung2025stellarmassdispersionmeasurecorrelations}.

To select an FRB sample, we first start with all localized FRBs in the literature. We apply an initial cut of $z<0.2$, leaving 67 FRBs. We compile stellar mass $M_*$, $i$, $\tau_\mathrm{exc}$, $b_\perp$, and RM data, borrowing heavily from Table 1 of \cite{Li_2025}. In Figure \ref{fig:sample} we have plotted the rest frame $\mathrm{DM_{host}}=\mathrm{DM_{host,ISM}}+\mathrm{DM_{host,CGM}}$ of each FRB, determined by subtracting from $\mathrm{DM_{obs}}$ the DM component estimates described above, against each $\mathrm{DM_{host,ISM}}$ indicator. It is  clear that $\tau_\mathrm{exc}$, RM, and $b_\perp$ are correlated with $\mathrm{DM_{host,ISM}}$. $i$ does not appear to correlate strongly by eye, which aligns with the relatively weaker correlation found by \cite{Li_2025}. With this information, we select five FRBs that we deem likely to have little $\mathrm{DM_{host,ISM}}$ contribution for our analysis, meeting the criteria of having two or more of the four indicators: $i<50$\textdegree, $\tau_\mathrm{exc}<1\,$ms, $|\mathrm{RM}|<100\,$rad m$^{-2}$, or $b_\perp>10\,$kpc. 
We select only from the FRBs whose localization is sufficient to determine a reliable $b_\perp$ with respect to the host galaxy. The FRBs and our reasoning for selecting them are described below, with notable properties summarized in Table \ref{tab:properties}. Figure \ref{fig:images} shows the FRB's host galaxies and localizations. Figure \ref{fig:pdfs} shows in detail the estimates of the DM components as described above for each FRB.

\begin{table*}[]
    \setlength{\tabcolsep}{12pt}
    \centering
    \begin{tabular}{lccccc}
        \hline 
        FRB & 20220319D & 20240209A & 20210807D & 20200120E & 20190425A \\
        \hline
        RA & 32.17792 & 289.88750 & 299.22143	& 149.47783 & 255.67500 \\
        DEC & 71.03526 & 86.06444 & -0.76236 & 	68.81889 & 21.57639 \\
        DM$_\mathrm{obs}$ & 110.95 & 176.52 & 251.30 & 87.82 & 128.20 \\
        $z$ & 0.0112 & 0.1384 & 0.1293 & 0.0008 & 0.0312 \\
        log$_{10}M_*/M_\odot$ & 9.93 & 11.34 & 10.94 & 10.86 & 10.26 \\
        log$_{10}M_\mathrm{h}/M_\odot$ \{vir,500c\} & 11.6, 11.5 & 13.2, 13.0 & 12.4, 12.3 & 12.3, 12.1 & 11.8, 11.6 \\
        $R_\mathrm{h}$ (kpc) \{vir,500c\} & 195, 102 & 615, 323 & 342, 182 & 326, 168 & 223, 117 \\
        $b_\perp$ (kpc) & $2.2^{+0.28}_{-0.27}$ & $39.7^{+3.5}_{-3.3}$ & $12.4^{+1.0}_{-1.0}$ & $20.1^{+0.0}_{-0.0}$ & $27.0^{+7.8}_{-7.8}$ \\
        References & \hyperlink{cap:note1}{1}, \hyperlink{cap:note2}{2} & \hyperlink{cap:note3}{3}, \hyperlink{cap:note4}{4} & \hyperlink{cap:note5}{5}, \hyperlink{cap:note6}{6} & \hyperlink{cap:note7}{7}, \hyperlink{cap:note8}{8} & \hyperlink{cap:note9}{9}\\
        \hline
    \end{tabular}
    \caption{FRB and host galaxy properties. $M_\mathrm{h,vir}$ is inferred from the stellar mass-halo mass relation with an associated uncertainty of 0.2\,dex \citep{Behroozi2019}. 
      \hypertarget{cap:note1}{}1.~\cite{Law_2024}
      \hypertarget{cap:note2}{}2.~\cite{Ravi_2025}
      \hypertarget{cap:note3}{}3.~\cite{Shah_2025}
      \hypertarget{cap:note4}{}4.~\cite{Eftekhari_2025}
      \hypertarget{cap:note5}{}5.~\cite{Shannon_2025}
      \hypertarget{cap:note6}{}6.~\cite{Gordon2023}
      \hypertarget{cap:note7}{}7.~\cite{Bhardwaj_2021}
      \hypertarget{cap:note8}{}8.~\cite{Kirsten_2022}
      \hypertarget{cap:note9}{}9.~\cite{bhardwaj2023hostgalaxiesnearbychimefrb}
    }
    \label{tab:properties}
\end{table*}

\textit{FRB 20220319D --} Reported in \cite{Law_2024} and analyzed in \cite{Ravi_2025}, FRB 20220319D has one of the lowest $\mathrm{DM_{obs}}$ of any FRB. This was exploited in \cite{Ravi_2025} to place a tight constraint on the CGM of the MW. This FRB has a very small $\tau_\mathrm{exc}<0.1\,$ms, low $|\mathrm{RM}|=59\,$rad m$^{-2}$, and the host galaxy is a nearly face-on spiral $i=24.6$\textdegree, making it an excellent candidate for our sample. 

\textit{FRB 20240209A --} This is the only known FRB originating from an isolated massive elliptical galaxy \citep{Shah_2025,Eftekhari_2025}. $b_\perp=40\,$kpc is 5 times the half-light radius of the galaxy in the Gemini r-band, meaning that the FRB likely comes from a globular cluster and has almost no $\mathrm{DM_{host,ISM}}$. $\tau_\mathrm{exc}<0.18\,$ms, aligning with this idea. FRB 20240209A is also an attractive target for our study because we expect AGN feedback to alter the CGM in halos of this mass \citep{McCarthy2010, Schawinski2007}, which is discussed in Section \ref{sec:elliptical}. 

\textit{FRB 20210807D --} This FRB has a large $b_\perp=11.91\,$kpc and low $i=44.5$\textdegree \citep{Gordon2023,Shannon_2025}. We include it in our sample with the caveat that it appears to have a non-negligible scattering (Figure B1 in \cite{Shannon_2025}). 
This means that a measurement of its CGM is more likely an upper limit (Section \ref{sec:analysis}).

\textit{FRB 20200120E --} A very nearby FRB originating from a globular cluster in M81 \citep{Bhardwaj_2021,Kirsten_2022}. This is the lowest $\mathrm{DM_{obs}}$ FRB at the time of writing, and an excellent candidate for our sample because it will have encountered little host ISM gas. Accordingly, it has a low RM and $\tau_\mathrm{exc}$. M81 is known to have an extended HI disk; \cite{Bhardwaj_2021} use a scaling relation between DM and HI column density to estimate $\mathrm{DM_{HI}}\sim5-10\,$pc/cm$^3$. While much of this HI could be classified as CGM, this effectively sets an upper limit on $\mathrm{DM_{host,ISM}}$ that validates our selection for low $\mathrm{DM_{host,ISM}}$ in this case.

\textit{FRB 20190425A --} This CHIME FRB's host galaxy is sufficiently nearby such that the $b_\perp$ is not terribly uncertain even without arcsec localization \citep{bhardwaj2023hostgalaxiesnearbychimefrb}. It is included in our sample because of the large $b_\perp=26.13\,$kpc and low $\tau_\mathrm{exc}=0.38\,$ms. We note that the host associations in \citep{bhardwaj2023hostgalaxiesnearbychimefrb} are made using strong priors from low measured DMs, which may bias a measurement of $\mathrm{DM_{host,CGM}}$. However, we are not performing a global population analysis in this work and our conclusions are not strongly dependent on a low $\mathrm{DM_{host,CGM}}$ for this FRB.

We note that several unaccounted for sources of DM could make the constraints on $\mathrm{DM_{host,CGM}}$ tighter. For instance, other bodies of gas such as a Local Group medium, M31, or the Magellenic clouds, could contribute tens of pc cm$^{-3}$ to the observed DM \citep{ProchaskaZheng2019}. M81 actually resides in a galaxy group, and FRB 20200120E passes within 60 kpc of M82 (halo mass roughly half of M81 \citep{Bhardwaj_2021}). If we included M82 in our analysis, the measured $\mathrm{DM_{host,CGM}}$ of M81 would decrease accordingly. Other FRBs in our sample and future samples may also have companion or satellite galaxies. Finally, our $\mathrm{DM_{MW,ISM}}$ estimates are lower than the NE2001 or YMW16 models, which are widely adopted for FRB science, and recent work indicates that $\mathrm{DM_{MW,CGM}}$ may exceed the predictions of the YT20 model (Section \ref{sec:mw}).

\hypertarget{textlink}{}
In Section \ref{sec::fgas} we include a sixth FRB in our analysis with a galaxy cluster host, to demonstrate the range of halo masses accessible with FRBs. This is the local universe cluster FRB 20220509G from \cite{Connor_2023}. In this case the interpretation is different than for our main sample because we are now probing the intra-cluster medium (ICM). For this FRB, we perform the same analysis as for our main sample, but we further subtract a $\mathrm{DM_{host}}$ component which represents the gas associated with the subhalo/galaxy from which the FRB originates. For the $\mathrm{DM_{host}}$ PDF we use the best-fit lognormal from \cite{Connor2025}. This FRB is included for demonstration but does not drive the results.

\section{Analysis}
\label{sec:analysis}

Now that we have isolated $\mathrm{DM_{host,CGM}}$ for our sample (under the assumption of negligible $\mathrm{DM}_\mathrm{host,ISM}$), we can use this to study the gas in these halos. $\mathrm{DM_{host,CGM}}$ is an integral of the electron density through the halo and so depends on both the radial profile, through $b_\perp$, and the overall normalization (the total amount of gas in the halo). As a function of halo mass, both of these pieces are of scientific interest and encode information about feedback processes. Throughout this work, we use the stellar mass-halo mass relation from \citep{Behroozi2019} to convert the host galaxy stellar mass measurements. 

\subsection{Comparison to simulations}

\begin{figure*}
    \centering
    \includegraphics[width=0.7\linewidth]{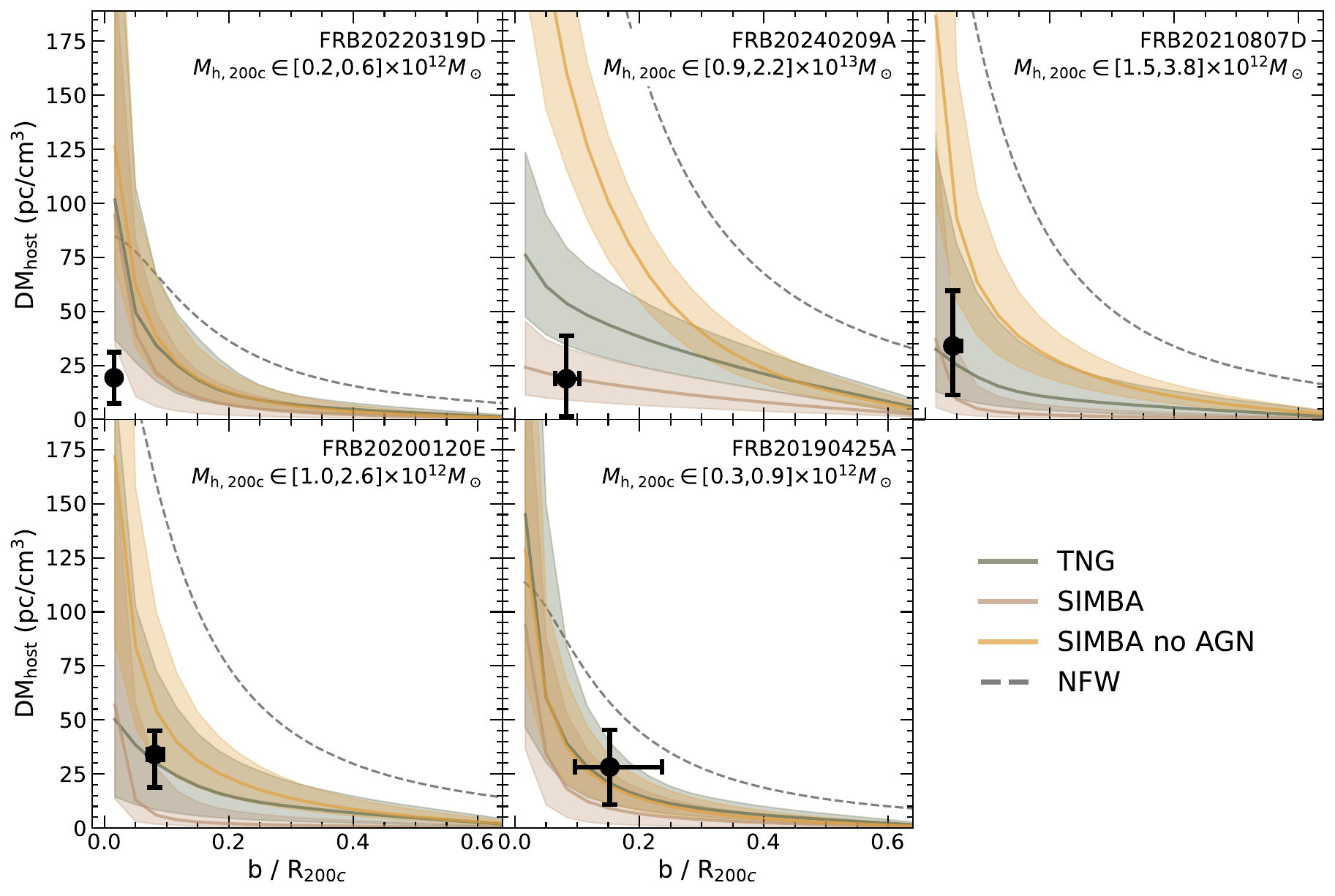}
    \caption{$\mathrm{DM_{host}}$ vs. normalized impact parameter through the halo $b_\perp/R_\mathrm{200c}$. Each panel shows the measurement from an FRB in our sample (in black) and randomly selected halos from simulations within 0.2\,dex of the FRB host $M_\mathrm{h,200c}$. We show three simulations: IllustrisTNG (TNG50) \citep{nelson2021illustristngsimulationspublicdata}, SIMBA \citep{Dav__2019}, and SIMBA with AGN feedback turned off. Solid lines are the median values, and shaded regions are the 16th and 84th percentile intervals, capturing both sightline-to-sightline variance within each halo and halo-to-halo variance. The x-axis uncertainty includes the FRB localization uncertainty and the uncertainty on host halo mass. The dashed grey curve is  the DM computed analytically for a halo gas profile exactly following the NFW profile at the nominal host halo mass with the cosmological average baryon fraction.}
    \label{fig:sims}
\end{figure*}

First, we compare our measurements to the halos seen in simulations. We use the IllustrisTNG (TNG) \citep{nelson2021illustristngsimulationspublicdata, Pillepich_2017,Springel_2017,Nelson_2017,Naiman_2018,Marinacci_2018} and SIMBA \citep{Dav__2019} simulations. Specifically, for TNG we use the TNG100-2 run, having a minimum particle gas mass of $1.1\times10^7\,M_\odot$. For SIMBA, we use the m100n1024 run with the fill s50 physics, and a gas mass resolution of $1.82\times10^7\,M_\odot$. We choose TNG100-2 over TNG100-1 because the resolution is closer to SIMBA, but verify that moving to TNG100-1 does not change the results significantly. We use the redshift 0 snapshot in all simulations. These two simulations are a good point of comparison because they will have different halo gas contents due to different AGN feedback implementations, especially at masses $\gtrsim10^{12}\,M_\odot$ \citep{Oppenheimer_2021}. We expect SIMBA to have gas-poor halos relative to TNG because of the stronger AGN feedback. SIMBA also provides versions of their simulation run with AGN feedback turned off (s50noagn), which we include in our analysis with the same box size. Although both the TNG and SIMBA simulations assume the Planck 2016 cosmology, we do not expect our results to be sensitive to small changes in the cosmology \citep{Planck_2016}. For each FRB and each simulation, we randomly select 100 halos uniformly in log space within 0.2dex of $M_\mathrm{h,200c}$ of the FRB host galaxy (the uncertainty in the \citep{Behroozi2019} stellar mass-halo mass relation). We use the {\tt yt} package to grid the simulation data uniformly in a sphere of radius $R_\mathrm{200c}$ around the center of the halo. We integrate the electron density along the positive and negative directions of each axis to get the DM, stopping at the midplane of the halo. Therefore, we have many sightlines for a given normalized $b_\perp/R_\mathrm{200c}$, capturing both the sightline-to-sightline variance within a halo and the variance between the 100 halos.  

We are assuming that the FRB is located exactly halfway through the halo along the line of sight. This is a good approximation for FRBs 20220319D, 20210807D, 20200120E, and 20190425A, where the FRB is very likely coming from the disk. For FRB 20240209A the situation is less clear. At an impact parameter of 40\,kpc, it is likely that FRB 20240209A originates from a globular cluster or satellite galaxy in its host. To quantify the likelihood that FRB 20240209A lies a given depth into the host halo, we use the distribution of globular clusters measured around M87 \citep{Harris_2009}. M87 is a large elliptical galaxy of comparable mass to the host of FRB 20240209A. The M87 system has thousands of detectable globular clusters due to its proximity. \cite{Harris_2009} find that the log of the globular cluster surface density in M87 decreases radially as $(r/R_\mathrm{eff})^{1/4}$, where $R_\mathrm{eff}$ is the effective radius of the galaxy. We make the assumption that this relation holds for the host of FRB 20240209A, and use it to determine that FRB 20240209A lies within $\sim\pm40\,$kpc of the midplane of the halo at $1\sigma$ uncertainty. Assuming reasonable CGM profiles (see the next section), this shift would change the measured $\mathrm{DM}_\mathrm{host,CGM}$ by up to $\sim$5 pc/cm$^3$ (25\% of the measured value). We include this as additional uncertainty on the measured $\mathrm{DM}_\mathrm{host,CGM}$ of FRB 20240209A.

The results are plotted in Figure \ref{fig:sims}. The uncertainty in $b_\perp/R_{200c}$ includes both the localization uncertainty and the uncertainty in $R_{200c}$ due to the halo mass. Note that below roughly $0.1R_\mathrm{200c}$ (10's of kpc) there will start to be contributions to the DM from the ISM in the simulated halos \citep{cordes2003ne2001inewmodelgalactic, Kravtsov_2013}, but we have not subtracted any $\mathrm{DM_{host,ISM}}$ component from the FRB data, so they are still comparable. For comparison, we also plot the expected DM profile for the scenario where baryons trace the underlying dark matter NFW profile exactly, with the cosmoligcal baryon fraction $\Omega_b/\Omega_m$ (see Section \ref{sec::fgas} for how this is computed).

Examining the case of FRB 20240209A, the data appear to favor the SIMBA simulation, but are also consistent with TNG within the uncertainty. On the other hand, the measurement definitively rules out the SIMBA no AGN feedback scenario. Without AGN feedback, halos of this mass ($\sim10^{13}\,M_\odot$) may retain too many of their baryons. Better agreement with SIMBA may also indicate that a stronger AGN feedback scenario is required than that implemented in TNG, although more data is needed to confirm this. Further, the data is inconsistent with the analytic NFW profile, which can be interpreted roughly as a no feedback scenario (although it also ignores other baryonic effects such as cooling).

We note that we have not yet made any assumptions about the host of FRB 20240209A, as we do in the following section (for example, spherical symmetry).
Furthermore, we note that in order to bring our data point of FRB 20240209A in agreement with the no feedback scenario, or a weaker than TNG scenario, we would require either: 1) we have significantly overestimated other DM components of FRB 20240209A. For example, $\mathrm{DM_{MW,ISM}}$ would need to be at least a factor of 2 smaller on this sightline to bring the data above the median of the TNG curve, which is unlikely by the nearby pulsar lower limit. 2) The host halo of FRB 20240209A is a significant outlier in terms of gas content. This in unlikely considering we have analyzed a representative population of 100 halos in each simulation. 3) FRB 20240209A is far in the foreground of the halo or not associated with the halo at all. The foreground of the halo scenario is unlikely as we discussed in the last paragraph, and the possibility that FRB 20240209A is not associated with the putative host galaxy is discussed in Section \ref{sec:hostassociation}.

The other four FRBs in the sample have less constraining power, but overall agree with simulations. FRB 20220319D marginally disagrees with all three simulations, but this may be explained by the ISM. For example, FRB 20220319D may be slightly in the foreground of its disk, leading the simulations to pick up relatively more $\mathrm{DM_{host,ISM}}$. Further, we specifically selected for low $\mathrm{DM_{host,ISM}}$ in the FRB host galaxies, but have not done so in the simulations. FRBs 20210807D, 20200120E, and 20190425A seem to align more closely with the TNG simulation over the SIMBA simulation, but the uncertainty is large enough to accommodate both. All four FRBs are inconsistent with the NFW profile scenario at various degrees of significance. These points are also higher overall than FRB 20240209A despite being at lower mass, which tentatively supports the idea that halos at $M_h\sim10^{12}\,M_\odot$ may retain a larger fraction of their baryons than more massive halos. A larger sample of local universe FRBs will provide a more stringent test of simulations and may be able to distinguish between the TNG and SIMBA simulations.

\subsection{The halo gas fraction}
\label{sec::fgas}

\begin{figure}
    \centering
    \includegraphics[width=0.8\linewidth]{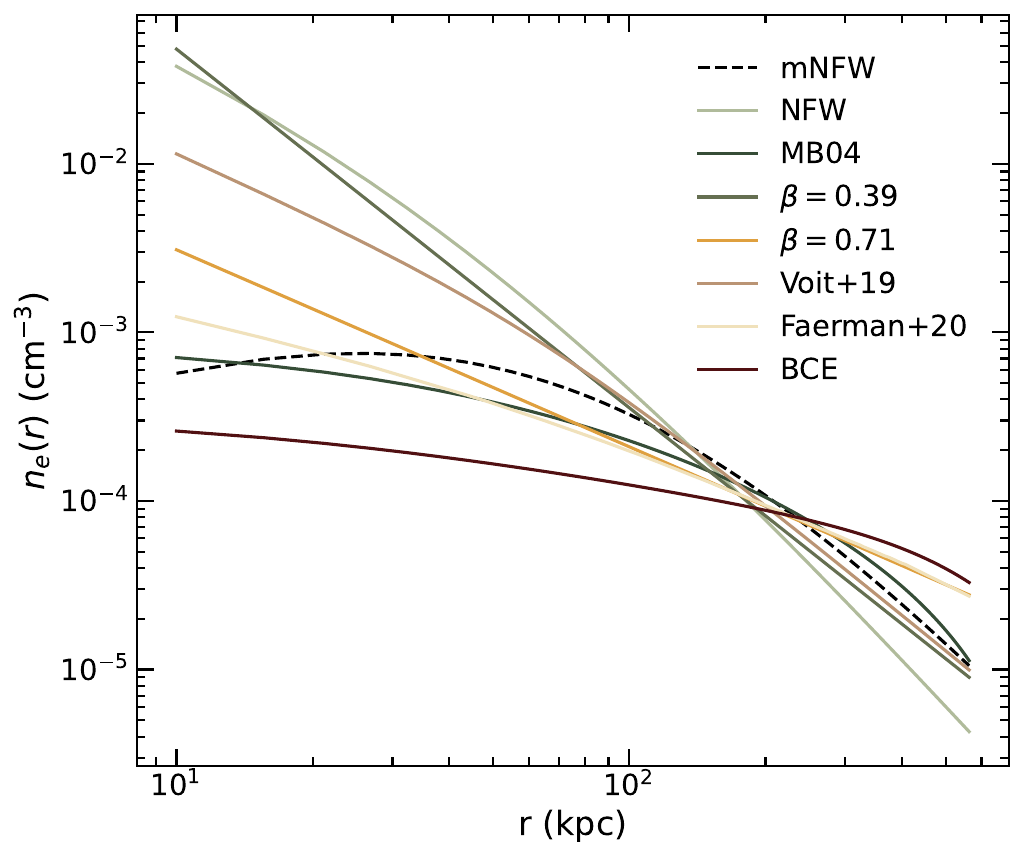}
    \caption{The CGM density profiles considered in our analysis, calculated for a galaxy with log$_{10}\,M_{h,\mathrm{vir}}=12.5$ and normalized to give 40 units of DM for an impact parameter of 20 kpc.}
    \label{fig:desnity_profiles}
\end{figure}

\begin{figure*}
    \centering
    \includegraphics[width=0.6\linewidth]{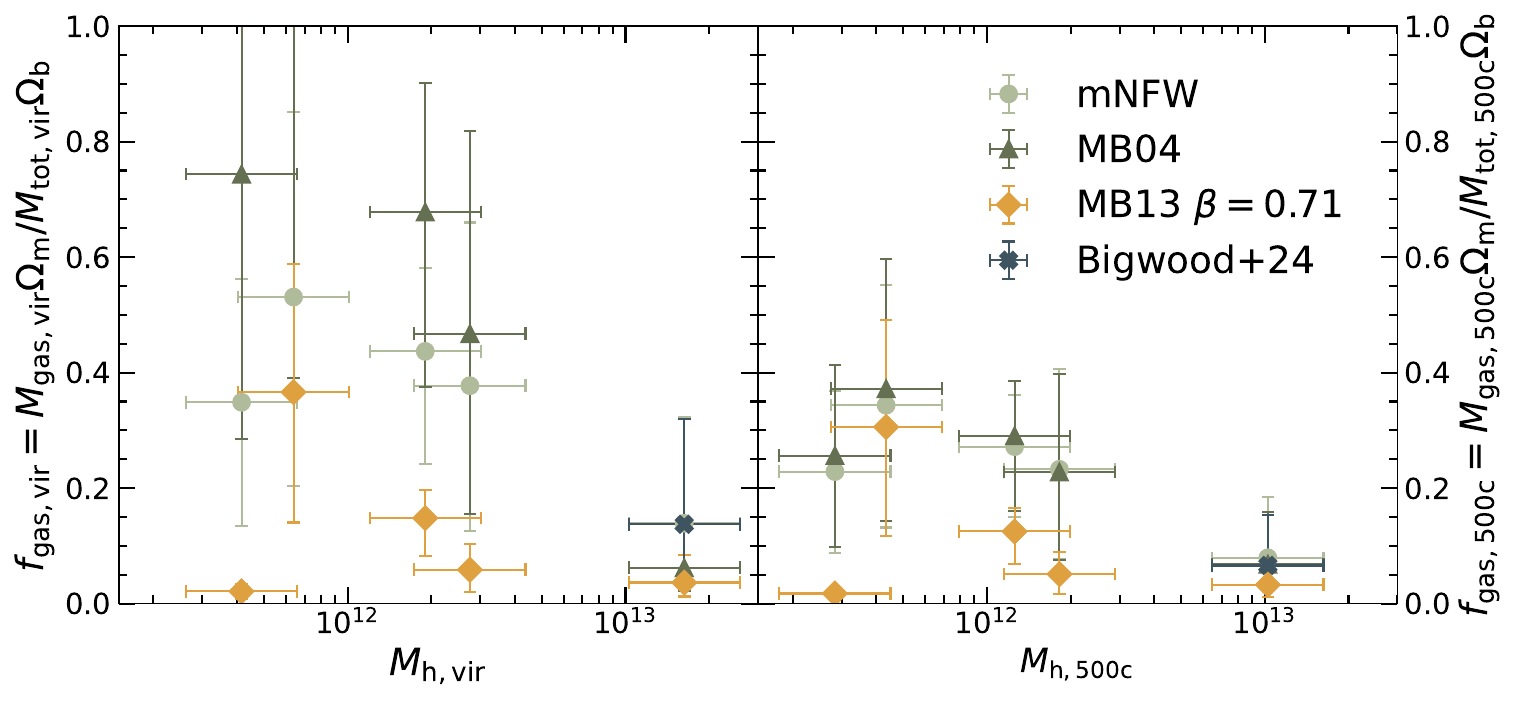}
    \caption{The halo gas fraction for our FRB sample plotted for a subset of the profiles considered. The left plot shows the gas fraction within the $R_\mathrm{vir}$, $f_\mathrm{gas,vir}$, and the right shows the gas fraction within $R_\mathrm{500c}$, $f_\mathrm{gas,500c}$. For clarity, we only plot our fiducial profile (mNFW), and the highest and lowest predicting profiles (excluding NFW). We also plot the \cite{bigwood2024weaklensingcombinedkinetic} profile for FRB 20240209A to show that it is consistent. Varying the radial profile between the several analytic forms considered introduces an uncertainty that is comparable to or smaller than the uncertainty from DM modeling. Our conclusions are robust to these changes. }
    \label{fig:profiles}
\end{figure*}

\begin{figure*}
    \centering
    \includegraphics[width=\textwidth]{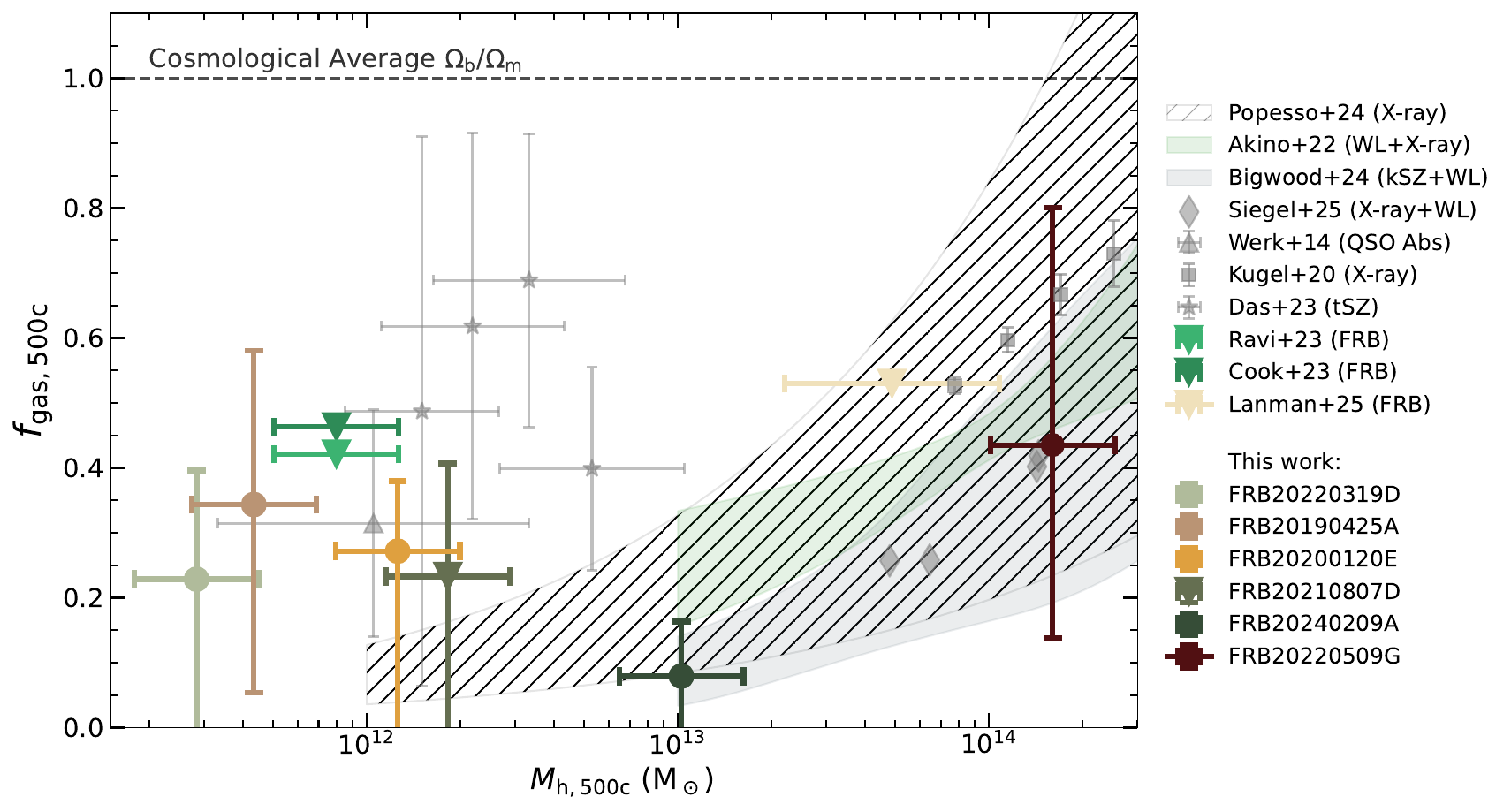}
    \caption{The halo gas fraction within $R_\mathrm{500c}$, $f_\mathrm{gas,500c}$, versus halo mass. The colored points are FRB data, including those analyzed in this work -- the five FRBs in our sample plus the cluster FRB 20220509G (see \hyperlink{textlink}{text}) -- and measurements from the literature. For comparison, we include representative studies from the literature that use complementary observational tools; see the text for details. Note that the data in the different studies represented here are not necessarily from the same redshift. }
    \label{fig:fgas}
\end{figure*}

We can take the analysis a step further and turn $\mathrm{DM_{host,CGM}}$ into a measurement of the gas content in the host halos. To translate the DM of a single FRB sightline through a halo into a constraint on the gas mass requires assumptions about the distribution of the gas. With a large sample of FRBs at various $b_\perp$ it would be possible to simultaneously fit the gas mass and the radial profile (Section \ref{sec:asm}), but this is limited by the amount of data at present. Instead, our approach is as follows: we assume that the gas distribution in a halo is well described by a spherically symmetric radial profile. We perform our analysis with various profiles proposed in the literature, and show that our main conclusions are largely robust to the choice of profile. We use the spread in predictions between these profiles as a proxy for the uncertainty introduced by our limited knowledge of the gas profile.

We continue our assumption that each FRB originates from halfway through the halo, except in the case of FRB 20240209A where we have added additional uncertainty to $\mathrm{DM_{host,CGM}}$ to account for the distribution of possible positions within the halo, as explained in the last section. Finally, we neglect the uncertainty in $b_\perp$ due to the localization of the FRBs, as only FRB 20190425A's localization uncertainty is large enough to have an impact on the analysis (and even then it is small). 

We largely follow \cite{ProchaskaZheng2019} and \cite{Keating_2020}. Given a radial density profile $\rho(r)$, normalized to the total halo mass $M_\mathrm{h,vir}$, and impact parameter $b_\perp$, the observed DM is then

\begin{equation}
\mathrm{DM}_\mathrm{host}
= \int\limits_{0}^{\sqrt{r_\mathrm{max}^2 - b_\perp^2}}
\rho(r)\, f_\mathrm{gas}\, \frac{\Omega_b}{\Omega_m}\,
\frac{\mu_e}{m_p \mu_H}\, ds
\label{eqn:dmhostintegral}
\end{equation}

\noindent where $s$ is the path length through the halo, and $r=\sqrt{s^2+b_\perp^2}$. We take $r_\mathrm{max}$ to be the halo virial radius $R_\mathrm{vir}$, and $\mu_e=1.167$, $\mu_H=1.33$ for an ionized mixture of hydrogen and helium.  $f_\mathrm{gas}=M_\mathrm{gas}\Omega_m/M_\mathrm{h}\Omega_b$ is the halo gas fraction relative to the cosmological average. Choosing $r_\mathrm{max}=R_\mathrm{vir}$ means we are measuring $f_\mathrm{gas,vir}=M_\mathrm{gas,vir}\Omega_m/M_\mathrm{h,vir}\Omega_b$. For our assumed cosmology, $\Omega_b/\Omega_m=0.1575$. 

The simplest assumption is that the halo gas follows the dark matter, well-modeled by the NFW profile \citep{NFW}:

\begin{equation}
    \rho(r)\propto\frac{1}{y(1+y)^2}
\end{equation}

\noindent where $y=r/r_s$, with the scale radius $r_s = R_\mathrm{vir}/C$, and $C$ the concentration parameter. For all dark matter halo calculations we use the COLOSSUS package \citep{Colossus}. We use the diemer19 model in COLOSSUS for the $M_\mathrm{h}-C$ relation so that the NFW is a single-parameter profile \citep{Diemer19}. The scatter in this relationship is appromixately 0.16 dex. The Here and in all following profiles the normalization is determined by requiring that the density profile integrates to $M_\mathrm{h,vir}$ at $R_\mathrm{vir}$.

The plain NFW profile is not a good representation of the gas in halos because we expect various baryonic effects to alter the distribution. \cite{MB04} proposes an alternative profile taking into account gas fragmentation:

\begin{equation}
    \rho(r)\propto\left[1+\frac{3.7}{y}\mathrm{ln}(1+y)-\frac{3.7}{C_c}\mathrm{ln}(1+C_c)\right]^{3/2}
\end{equation}

\noindent where $C_c=r_c/r_s$ and $r_c$ is a cooling radius. We follow \cite{ProchaskaZheng2019} and assume constant $r_c=147\,$kpc, which is a decent approximation given that it is a weak function of halo properties (it will change by a few tens of kpc over the halo masses of interest here, see Figure 3 in \cite{MB04}) and significantly larger than the impact parameters explored here, such that it shouldn't affect the results significantly. 

An alternative correction to the NFW, the modified NFW (mNFW), is presented by \cite{MathewsProchaska2017} and generalized by \cite{ProchaskaZheng2019} as:

\begin{equation}
\label{eqn:mnfw}
    \rho(r)\propto\frac{1}{y^{1-\alpha}(y_0+y)^{2+\alpha}}
\end{equation}

\noindent where $y_0=\alpha=2$ is typically chosen. Thus the inner slope of the CGM is actually rising with radius. \cite{MathewsProchaska2017} explore a range of $y_0$ from 1-4, with $y_0=2$ matching observations well. The range of $\alpha$ is not formally bounded, but realistically it could vary from 0 (the NFW profile) to $\sim3$, with values beyond that producing unphysical profiles.

\cite{Voit19} create a pNFW profile, which we implement from the provided tables. The assumption of this profile is that cooling and percipitation of the gas is regulated by feedback processes. The above three profiles decrease the density of the gas at small radii compared to the NFW and distribute it farther out, approximating baryonic effects like feedback.

We also include the $\beta$ model from \cite{MillerBregman13}, determined empirically from X-ray data:

\begin{equation}
    n_e(r)\propto\left(1+\left(\frac{r}{r_c}\right)^2\right)^{-3\beta/2}
\end{equation}

\noindent where $r_c=0.35\,$kpc. We assume $r_c$ scales with $r_s$ for different halos, but the total gas mass in the halo is insensitive to $r_c$ such that the exact choice is unimportant. As stated in \cite{Ravi_2025}, typical values for $\beta$ determined by aborption line and X-ray studies are $\sim0.4-0.5$ \citep{MillerBregman13, Miller_2015,Kaaret_2020, Zhang_2024, Popesso_2026}. We will evaluate this profile with both $\beta=0.71$ \citep{MillerBregman13} and $\beta=0.39$ \citep{Kaaret_2020}, which bracket the range of plausible values.

We further include the model of \cite{Faerman_2020}. We take the tabulated fiducial model for the MW and scale to the virial radius of other halos. The shape of the model in \cite{Faerman17} is similar and so not included separately. 

All of the above profiles are motivated by or fit to observations of the MW. Because the host of FRB20240209A is a massive elliptical, these may not generalize well. We include the ``baryonification" model \citep{Schneider_2019}, with best-fit parameters from the kinetic Sunyaev-Zeldovich effect (kSZ) study by \cite{bigwood2024weaklensingcombinedkinetic}. This model alters the NFW profile in detail to emulate the observed gas distributions in X-ray observations of massive halos. For our purposes, the gas profile is:

\begin{equation}
    \rho(r)\propto \frac{1}{\left[1+\left(\frac{r}{0.1R_{200c}}\right)\right]^{\beta(M)} \left[1+\left(\frac{r}{r_{ej}}\right)^\gamma\right]^{\frac{\delta-\beta(M)}{\gamma}}}
\end{equation}

\noindent where $\beta(M)=3(M/M_c)^\mu/(1+(M/M_c)^\mu)$ and $r_{ej}=\theta_{ej}R_{200c}$. $M_c$, $\theta_{ej}$, $\delta$, $\gamma$, and $\mu$ are the parameters retrieved from the best-fit by \cite{bigwood2024weaklensingcombinedkinetic}. 

The profiles described above are plotted in Figure \ref{fig:desnity_profiles} for an example galaxy and FRB. Given an observed DM$_\mathrm{host}$, $b$, $M_\mathrm{h,vir}$, and density profile, $f_\mathrm{gas,vir}$ is determined by Eqn. \ref{eqn:dmhostintegral}. Once $f_\mathrm{gas,vir}$ is determined for a halo, where we have assumed the DM probes out to the virial radius, we can then reintegrate the density profile to get $f_\mathrm{gas,500c}$, the gas mass fraction within $R_{500c}$, which we use to compare to other observations below. $f_\mathrm{gas,vir}$ and $f_\mathrm{gas,500c}$ determined for each FRB for a subset of the profiles described above are shown in Fig. \ref{fig:profiles}. There is significant scatter in $f_\mathrm{gas,vir}$, but this is reduced in $f_\mathrm{gas,500c}$. Overall, varying the radial profile introduces an uncertainty that is comparable to or smaller than the uncertainty from modeling the different DM components. Our main conclusions are largely robust to the choice of profile, see below. We take the mNFW profile as our fiducial prediction for all five hosts, because it is roughly the median prediction of the profiles for $f_\mathrm{gas,vir}$. Choosing this profile results in values of $f_\mathrm{gas,500c}$ that are on the high end of the different profiles, which is the conservative choice for our conclusions about FRB 20240209A. As explained previously, we include the scatter in the profile predictions around the mNFW value as additional uncertainty on our measurement. This approximates the uncertainty from our limited knowledge of the actual gas distribution. The FRB 20240209A measurement is largely insensitive to the choice of profile, and the mNFW profile is nearly identical to the \cite{bigwood2024weaklensingcombinedkinetic} profile, justifying our  choice of mNFW as our fiducial profile even in this higher-mass case.

Our fiducial measurements (mNFW profile) of $f_\mathrm{gas,500c}$ for the FRBs in our sample are plotted in Figure \ref{fig:fgas} and compared to the literature. FRB FRB20210807D is plotted as an upper limit because its non-negligible $\tau_\mathrm{exc}$ means there is likely a $\mathrm{DM_{host,ISM}}$ contribution (Section \ref{sec:sample}). We include the local universe cluster FRB 20220509G, discussed at the end of Section \ref{sec:sample}, in the plot to show that FRBs can probe this mass scale. The uncertainty on the mass of the corresponding cluster is 0.2dex \citep{Yang_2021}. For this FRB, we plot the fiducial prediction as the \cite{bigwood2024weaklensingcombinedkinetic} profile and do not include the uncertainty from other profile estimates. For FRB 20230703A from \cite{lanman2025constrainingbaryonfractionsgalaxy}, we have taken the value determined for the mNFW profile and the mass error is from the range of masses in the three groups along the FRB sightline. We plot this FRB as an upper limit because the authors did not consider $\mathrm{DM_{host}}$ or $\mathrm{DM_{cosmic}}$. The \cite{Cook_2023} and \cite{Ravi_2025} FRB points are upper limits on the MW CGM, for which we take the DM and convert it into an $f_\mathrm{gas}$ measurement using the same procedure for the FRBs in our sample. These points are plotted at the MW halo mass with 0.2dex uncertainty. Note that \cite{Ravi_2025} uses FRB 20220319D, which is also plotted separately for our sample. 

For the COS-Halos quasar absorption line study (QSO Abs) \citep{Werk_2014}, we have translated their limits on the cool mass of the CGM from $R_\mathrm{vir}$ to $R_\mathrm{500c}$ using their best fit surface mass profile. That is, they find that the fraction of cool gas in the CGM, $f_\mathrm{cool,vir}$ is between 0.25 and 0.45 for an average halo mass of $10^{12.2}\,M_\odot$. The hydrogen column surface density is $\propto r^{-1\pm0.5}$, meaning the gas mass scales as $r^{1\pm0.5}$, while the dark matter scales as the typical NFW profile. We can therefore extrapolate $f_\mathrm{cool,vir}$ to $f_\mathrm{cool,500c}$, accounting for the uncertainty in the profile. The galaxy sample has a scatter in stellar mass of approximately 0.5 dex, which roughly corresponds to the same in halo mass. Note that because this study traces only the cool phase of the CGM it is plotted as a lower limit. For \cite{Das_2023}, we convert their $M_\mathrm{gas,200c}$ measurement for stellar mass bins of width 0.3dex to $M_\mathrm{gas,500c}$ using their generalized NFW profile. Because their profile is fixed, this does not amplify the uncertainty. We report the halo mass error as the range of halo masses corresponding to their stellar mass bins. For \cite{popesso2024hotgasmassfraction} we use their best fit $f_\mathrm{gas,500c}$ power-law, and for \cite{Akino22} we use the best fit evaluated at $z=0$, in both cases using the quoted uncertainties. For \cite{bigwood2024weaklensingcombinedkinetic} and \cite{siegel2025jointxraykineticsunyaevzeldovich} we take their $M_\mathrm{gas,500c}$ data and convert using our assumed cosmology. \cite{Kugel23} is plotted as given.

As seen in Figure \ref{fig:fgas}, FRBs can constrain the gas content in a wide range of halo masses. We compare FRBs as a probe of halo gas to other observational tools in Section \ref{sec:otherprobes}. The FRB 20240209A measurement is the most constraining, and low compared to other observations. We interpret this as evidence for strong AGN feedback in halos of this mass, and discuss this in Sections \ref{sec:otherprobes} and \ref{sec:elliptical}.

\subsection{Gas radial profile}
\label{sec:mcmc}

\begin{figure}
    \centering
    \includegraphics[width=\linewidth]{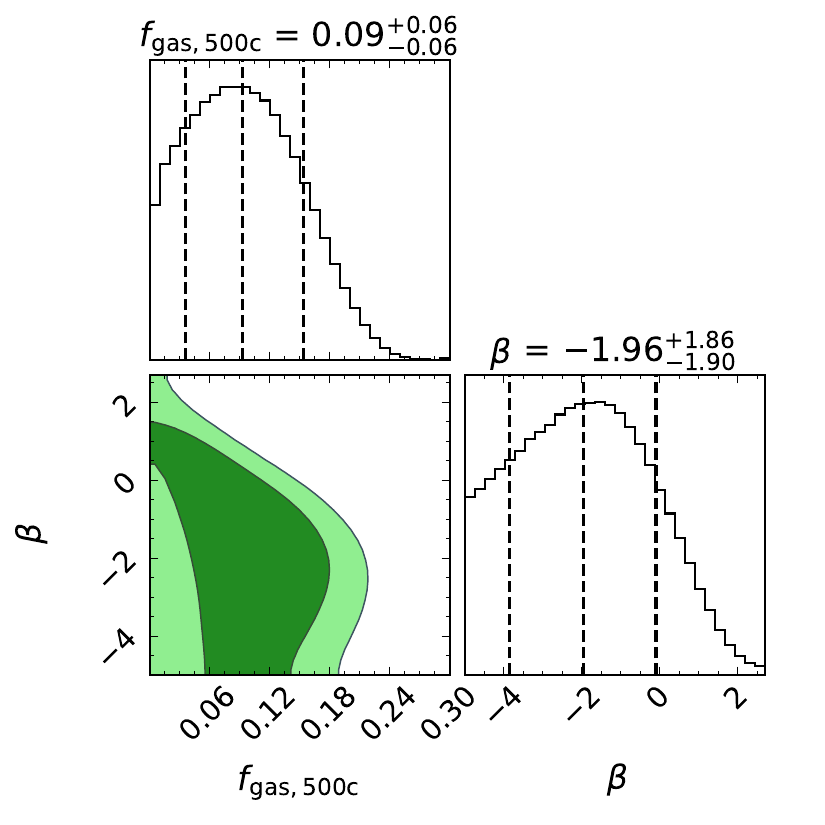}
    \caption{Corner plot of the MCMC fit to our FRB sample. $\alpha$ is the inner power-law slope of the CGM radial profile. The contours represent the 1 and 2$\sigma$ confidence intervals. The sample is not large enough for a good constraint on $\alpha$, but the fit demonstrates that there is constraining power in the data that can be taken advantage of with a future larger sample.}
    \label{fig:corner}
\end{figure}

The FRB data also encodes information about the radial profile of the CGM through the different $b_\perp$ of the FRBs. While the DM is determined by an integral of the electron profile out to $R_\mathrm{vir}$, sightlines with different $b_\perp$ give a differential measurement of the profile at small radii. Feedback determines the radial profile; stronger feedback will push baryons farther out of the halo, producing a flatter profile. 
With a large sample (Section \ref{sec:asm}), it will be possible to fit both the normalization, $f_\mathrm{gas}$, and the shape of the CGM, ideally in mass bins.
To demonstrate this, we use Markov Chain Monte Carlo (MCMC) to fit a global $f_\mathrm{gas,500c}$ and inner CGM power law slope $\alpha$ to our sample of five FRBs. Our sample is small and may be biased by individual FRBs or our sample selection; however, this analysis serves to demonstrate the methodology.

For this analysis, we re-parameterize the mNFW radial profile (Equation \ref{eqn:mnfw}) as 

\begin{equation}
    \rho(r)\propto\frac{1}{y^{\beta}(y_0+y)^{3-\beta}},
\end{equation}

\noindent so that $\beta$ represents the inner slope of the profile and the profile converges to the standard NFW at large radii where we do not have data. Here $y=r/r_s$ as before, where $r_s$ is the scale radius of the NFW, and the profile is normalized to the mass of the host halo. We again assume spherical symmetry and that the FRBs originate from halfway throught the halo along the line of sight. We chose $y_0=2$ as in the mNFW profile, so that all of the FRBs are originating from radii smaller than the transition between the inner and outer slopes. We use uniform priors on both parameters, with ranges $\beta \in [-5,5]$ and $f_\mathrm{gas,500c}\in [0.0,0.5]$. The likelihood for each host is the probability of the model predicted DM along the sightline given by the measured $p(\mathrm{DM}_\mathrm{host,CGM})$ (Figure \ref{fig:pdfs}).  The results of the fit are shown in Figure \ref{fig:corner}. Clearly a larger sample is necessary for a good determination of $\beta$, but the fit demonstrates that there is constraining power in the data. Interestingly, the data prefer a rising inner CGM profile, but a larger sample is required to verify this. The best fit $f_\mathrm{gas,500c}$ is likely heavily influenced by FRB 20240209A, which we expect to give a much tighter constraint than the other FRBs (Figure \ref{fig:fgas}), and is not 1-1 comparable to our other measurements because of the different profile. 

\subsection{$f_\mathrm{CGM}$}

Once we have determined $f_\mathrm{gas}$ for a range of halo masses, it is interesting to consider what this implies about the cosmological baryon budget. In particular, we can use our measurements to determine $f_\mathrm{CGM}\equiv\Omega_\mathrm{CGM}/\Omega_\mathrm{b}$ by integrating over the halo mass function (MF). Recently, \cite{Connor2025} provided a full accounting of the Universe's baryons, including the ``missing" baryons of the IGM and CGM. They used FRBs to determine $f_\mathrm{IGM}$, but $f_\mathrm{CGM}$ was determined indirectly by constraining other baryonic components (e.g. the ICM with X-rays) and requiring the sum of all components to reach $\Omega_\mathrm{b}$. They found $f_\mathrm{CGM}=0.08^{+0.07}_{-0.06}$ for halos of mass $10^9-5\times10^{12}\,M_\odot$. Using the tinker08 MF in COLOSSUS \citep{Tinker2008} and a weighted average of $f_\mathrm{gas,vir}$ for our 4 FRBs within the same mass range, we find $f_\mathrm{CGM}=0.11\pm0.07$. Thus we extend the ability of FRBs to account for the missing baryons beyond just the IGM. We have seen that FRBs can also constrain gas in the group and cluster scale halos (Figure \ref{fig:fgas}), so in principle FRBs alone can account for all components of the diffuse ionized gas in the Universe.

\section{Discussion}
\label{sec:discussion}

\subsection{Comparison to other probes}
\label{sec:otherprobes}

Because the DM is a simple integral of the electron density $n_e$, FRBs do not suffer from some of the limitations of other gas probes. This comes at the cost of a single line of sight through a halo per FRB. We have shown that FRBs can measure gas in halos of mass $\sim10^{11-13}\,M_\odot$ (Section \ref{sec::fgas}). While FRBs have been shown to preferentially occur in massive star-forming galaxies \citep{Sharma2024}, at least three known FRBs have dwarf host galaxies \citep{Tendulkar_2017,Bhandari_2020, Hewitt_2024}. Similarly, the existence of a few FRBs with massive cluster hosts \citep{Connor_2023,frbcollaboration2025cataloglocaluniversefast,lanman2025constrainingbaryonfractionsgalaxy}, including FRB 20220509G, indicates that future FRBs could be discovered in the most massive systems (up to $\sim10^{15}\,M_\odot$).  So a large sample of localized FRBs can in principle constrain the gas content of halos of all sizes. 

This is in contrast to the other probes (the k/tSZ effect, aborption line studies, and X-rays) which can be seen in the representative works that we have plotted in Figure \ref{fig:fgas}. The tSZ effect is $\propto n_eT_e$, both of which depend broadly on the halo mass, so that it is difficult to detect for low to intermediate mass halos. The state-of-the-art measurements are highly uncertain for MW mass halos and below \citep{Das_2023,Das_2025}. The kSZ signal also depends on the halo mass ($\propto n_e v$, where $v$ is set by large-scale structure), and the signal is intrinsically faint such that it can only be measured down to $10^{13}\,M_\odot$ with current instruments \citep{bigwood2024weaklensingcombinedkinetic}. The quasar absorption line studies probe only the cool and warm phases of the CGM ($\lesssim10^6\,$K), and so cannot in general study the gas of large groups and clusters \citep{Tumlinson_2017}. Finally, the X-ray signal ($\propto n_e^2$) is also strongest in the most massive systems. Recent eROSITA results have pushed down to $\sim 10^{12.5}\,M_\odot$ \citep{popesso2024hotgasmassfraction, Zhang_2024}. 

Another interesting detail to note is that we have shown that FRBs can measure individual halos. The kSZ effect, for example, is measured statistically. Similarly, FRB analyses up to this point have focused on statistical measurements of baryonic feedback (e.g. the $p(\mathrm{DM}|z)$ relationship \cite{reischke2025measurementbaryonicfeedbackfast}). This makes our method interesting on a per-source astrophysical level and a good complement to other FRB analyses. On the other hand, because we have only measured the gas in a handful of individual halos, we need a larger sample to make definitive claims. We can expect a larger sample in the near future (Section \ref{sec:asm}). 

Because of the small sample size and the relatively large errors (Figure \ref{fig:fgas}), we cannot make broad conclusions when comparing the FRB data to the literature. For FRBs 20200120E and 20210807D, the errors are large enough to be consistent with all of the tSZ, X-ray, and absorption line data \citep{Das_2023,popesso2024hotgasmassfraction,Werk_2014}. However, the FRB 20240209A measurement is less uncertain and quite low. It appears to be slightly in tension with both the pre-eROSITA X-ray measurements \citep{Akino22} and possibly the high-mass end of the tSZ sample from \citep{Das_2023}. We note that the measurement is consistent with the eROSITA X-ray measurements \citep{popesso2024hotgasmassfraction,Zhang_2024}.

A small gas fraction at the group scale implies stronger baryonic feedback and a stronger suppression of the matter power spectrum. This is expected in a stronger AGN feedback scenario, which we discuss in the next section. When taken together, the four low mass measurements (FRBs 20210807D, 20200120E, and 20190425A) indicate that $f_\mathrm{gas}$ may break from the decreasing trend that is seen when moving down in mass from $M_h\sim10^{15}\,M_\odot$ to $\sim10^{12.5}\,M_\odot$, which is independent of our choice of gas profile. This is in agreement with other observables (\cite{Werk_2014}, \cite{Das_2023}) and the interpretation that $L_*$ galaxies retain a larger fraction of their bayons than massive galaxies and galaxy groups due to the prevalence of AGN feedback in the latter. More data is required to confirm this trend.

\subsection{FRB 20240209A - AGN feedback}
\label{sec:elliptical}

\begin{figure}
    \centering
    \includegraphics[width=\linewidth]{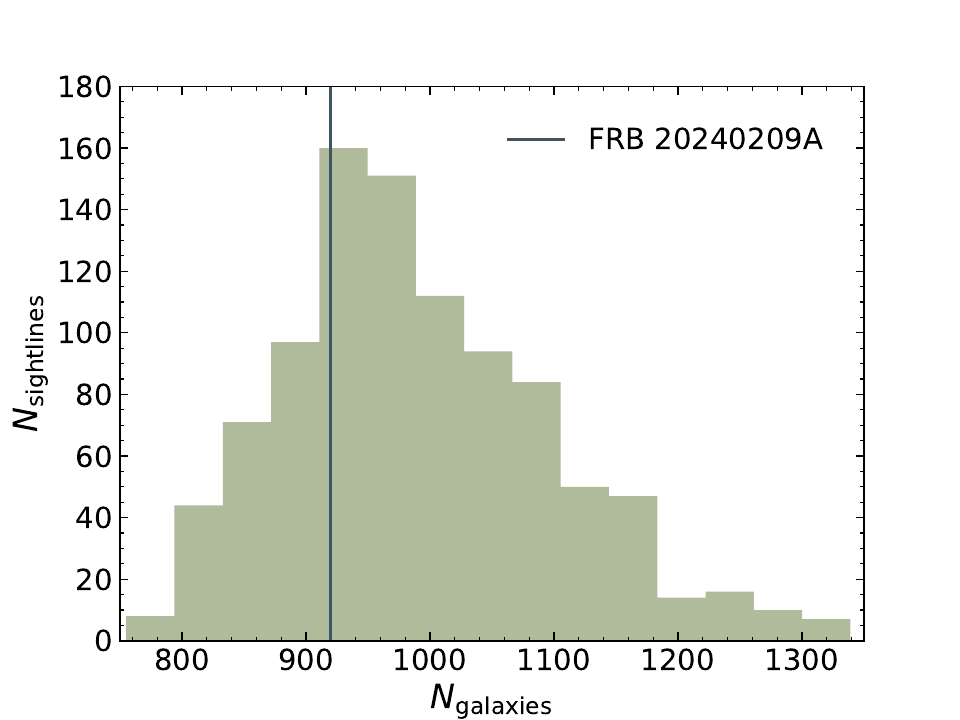}
    \caption{The number of galaxies within 5 Mpc of the sightline to FRB 20240209A compared to the distribution for 1000 random sightlines following the procedure of \cite{Hussaini_2025}. The FRB 20240209A sightline is consistent with the median overdensity of foreground galaxies to this redshift.}
    \label{fig:sightlines}
\end{figure}

The FRB data indicates that there is little gas in the host of FRB 20240209A. It is expected that halos of mass $\sim10^{13}\,M_\odot$ will be strongly affected by AGN feedback, which can push baryons out to the virial radius or beyond and prevent further accretion \citep{McCarthy2010,Oppenheimer_2021}. This process creates the quenched elliptical galaxies we see in the local universe, such as the host of FRB 20240209A \citep{Springel2005}. Our measurement of $\mathrm{DM_{host,CGM}}$ in the host of FRB 20240209A aligns most closely with the SIMBA simulation (Figure \ref{fig:sims}), which has a stronger AGN feedback implementation than IllustrisTNG, but is consistent with both within the uncertainty. It is further completely inconsistent with the SIMBA simulation with AGN feedback turned off. When converted to a measurement of $f_\mathrm{gas}$ (Figure \ref{fig:fgas}), it is most consistent with the findings of \cite{bigwood2024weaklensingcombinedkinetic}, which favor a stronger AGN feedback scenario. Below we discuss further evidence in support of the interpretation that AGN feedback has evacuated the host of FRB 20240209A. 

First, we note that the low $\mathrm{DM_{host,CGM}}$ could be explained by sightline-to-sightline variance in the DM components. While this should be captured in our uncertainty estimates, we check that the sightline to FRB 20240209A is not underdense, leading to a low $\mathrm{DM_{cosmic}}$. To do this we follow the procedure of \cite{Hussaini_2025}. We use the photometric redshift catalog of \cite{Beck2022}, which combines Pan-STARRS and Wide-field Infrared Survey Explorer (WISE) data. We count the number of galaxies in a 5 Mpc cylinder towards FRB 20240209A and 1000 random sightlines within 5 degrees of galactic latitude. The number of galaxies in each sightline is plotted in Figure \ref{fig:sightlines}. The sightline to FRB 20240209A is only slightly underdense, so this is unlikely to explain the low $\mathrm{DM_{host,CGM}}$ measurement. On the other hand, we have not accounted for sightline-to-sightline variance in the CGM of the host galaxies themselves when converting $\mathrm{DM_{host,CGM}}$ to $f_\mathrm{gas}$. This variance is captured in the curves calculated for the simulation halos (Figure \ref{fig:sims}), and appears to be roughly 50\% including halo-to-halo variance. This is similar to the 0.2dex uncertainty we assumed for $\mathrm{DM_{MW,CGM}}$. This uncertainty is not enough to significantly change our conclusions. Finally we check the local environment around the host of FRB 20240209A, as central galaxies in groups or clusters often host AGN: we find only a minor overdensity within 5 Mpc around the host compared to the average density in the catalog at this redshift, and only one galaxy closer than 3 Mpc, so the galaxy is likely not part of a group.

Next, we consider the available information about the host of FRB 20240209A from multiwavelength data. The host is analyzed in detail in \cite{Eftekhari_2025}. Using a deep Gemini spectrum and SED fitting, they find strong evidence that the host is a quenched elliptical. In particular, the very low star formation rate (SFR) ($<0.36\,M_\odot\,\mathrm{yr}^{-1}$), which is apparent in the lack of emission lines in the spectrum, combined with the high stellar mass places the host securely below the star forming main sequence (the specific SFR (sSFR) is log$_{10}$sSFR $<-11.8$). The stellar population age inferred from SED fitting is $t_m\sim11\,$Gyr, and indicates most of the stellar mass was formed around this time. Further, the host is squarely in the elliptical region of the WISE color diagram. Combined with the very low gas content of the halo inferred from the FRB in this work, the obvious interpretation is that this galaxy has been quenched by a period of strong AGN feedback after $z\sim2$. However, the WISE colors do not classify the host as an AGN, indicating the lack of a dusty torus. There is also a lack of any obvious optical emission lines associated with AGN activity. This seems to indicate that the nuclear supermassive black hole is not currently active or in a low power state.

Another piece of information that was not examined in detail by \cite{Eftekhari_2025} is the radio detection of the host by \cite{Law2024Atel}. An unresolved source coincident with the host was observed at 1-2 GHz with the VLA. The resolution is not sufficient to definitively associate the radio source with the nucleus of the host galaxy. However, if we assume that the radio emission in the host is entirely synchrotron from star formation, using the radio luminosity-SFR relation gives a SFR $\sim5\,M_\odot\mathrm{yr}^{-1}$ \citep{Murphy_2011}. This is an order of magnitude above the SFR inferred from the optical and infrared data, implying that there is a significant contribution to the radio emission from an AGN. 
A simple power-law fit to the flux densities gives a spectral index of $\alpha=-1.84$ ($S\propto\nu^\alpha$, where $S$ is the flux density and $\nu$ is frequency). This is steeper than expected from either star formation or an active radio-loud AGN.
The steep spectral index may be consistent with an aged synchotron population, for example if a past AGN has become dormant and is now producing less relativistic electrons. 

Based on the above discussion, we conclude that our measurement of a small $f_\mathrm{gas}$ in the host of FRB 20240209A is likely due to AGN feedback in the host galaxy that quenched star formation and expelled gas. The AGN may have since turned off or entered a low-power state. Thus we show that FRBs can measure feedback in individual halos, opening a new pathway to understanding baryonic effects. 

\subsection{FRB 20240209A - Host galaxy association}
\label{sec:hostassociation}
FRB\,20240209A is significantly offset from its host elliptical galaxy \citep{Shah_2025}. It is therefore worth asking if the FRB is associated 
with the putative host. The FRB is unlikely to be in a background galaxy, due to the asymmetry of cosmic DM \citep{Connor2025} and the tight total DM budget of this source. If it were, the CGM constraints would be even more significant because it would have more IGM DM and double the path length through the host galaxy's halo. Thus, the failure mode of our analysis is if the FRB is actually in the foreground of this galaxy. This is unlikely given the depth of optical imaging at the position of the FRB and 
the relative low redshift of the elliptical. No galaxy has been detected down to $r\approx25.9$\,mag (3-$\sigma$), ruling out galaxies with luminosity above $5\times10^6$\,$L_\odot$ that are halfway to the elliptical host candidate. Furthermore, elliptical galaxies are known to have far more globular clusters than late-type galaxies of similar mass. FRB\,20200120E, which is also in our sample, resides in a globular cluster of M81, providing precedent for large host offsets. As in the discovery papers \citep{Shah_2025,Eftekhari_2025}, we conclude that FRB\,20240209A is indeed in the outskirts of the massive elliptical, whether in a globular cluster or offset due to the progenitor's advanced stellar age.

\subsection{Host ISM and CSM}
\label{sec:hostism}

We have assumed throughout that host DM from the 
ISM and circumsource material (CSM) is negligible for our sample. Indeed, we have selected for local Universe FRBs with indicators of 
limited non-CGM host contribution, such as low scattering, low RM, large impact parameter, and face-on orientations. However, without concrete methods to quantify $\mathrm{DM_{host,ISM}}$ in distant galaxies, it is always possible that the host ISM still contributes to the observed DM. $\mathrm{DM_{host,ISM}}$ will dominate over the DM error budget from other components if it is greater than a few 10's of pc cm$^{-3}$, depending on the source (Figure \ref{fig:pdfs}). This is not a concern for our anaylsis because any residual 
$\mathrm{DM_{host,ISM}}$ contributions will drive our measurements of $\mathrm{DM_{host,CGM}}$, and thus $f_\mathrm{gas}$, even lower. This in turn puts further pressure on feedback scenarios in which galaxies retain a large portion of their baryons.  Figure's~\ref{fig:sims} and~\ref{fig:fgas} show 
that most of the constraining power of our sample already 
comes from the upper-bound on $\mathrm{DM_{host}}$, with $f_\mathrm{gas}$ values nearly consistent with zero. Our conclusion about strong AGN feedback in the host of FRB 20240209A is driven by the small measurement of $\mathrm{DM_{host,CGM}}$ and $f_\mathrm{gas}$. A larger future sample of FRBs
could produce a more meaningful bounded constraint on host ISM and CSM by better calibrating inclination angle, RM, scattering, and impact parameter to $\mathrm{DM_{host}}$.

\subsection{The Milky Way Halo}
\label{sec:mw}

One of the more uncertain aspects of our analysis is $\mathrm{DM_{MW,CGM}}$. $\mathrm{DM_{MW,CGM}}$ is dominated by a hot volume-filling gas phase, which is currently best measured in the X-ray \citep{Das2021}. In principle, OVII measurements correlate with the DM \citep{ProchaskaZheng2019, Das2021}, which could allow independent constraints on this component. However, the uncertainty on OVII column measurements from X-ray absorption is large, and the number of sightlines is limited (e.g. \cite{Fang_2015}). The YT20 model fits X-ray EM data; however, the model includes a spherical and ``disk"-like component, only the latter of which is fit to the data \citep{Yamasaki_2020}. The spherical component assumes an $f_\mathrm{gas}$ for the MW, and dominates high galactic latitude sightlines. Further, there is likely significant sightline-to-sightline variance in $\mathrm{DM_{MW,CGM}}$, given the large scatter in the measurements of both \cite{Das2021} and \cite{Yamasaki_2020}. Finally, there are a large number of models in the literature for the halo gas distribution, which produce widely varying predictions for $\mathrm{DM_{MW,CGM}}$ \citep{ProchaskaZheng2019, Keating_2020}. 

We wish to be conservative for our estimate of $\mathrm{DM_{MW,CGM}}$. \cite{Das2021} provides a comprehensive compilation of data on all phases of the MW CGM and converts these to DM estimates. These estimates are on average twice as high as those from YT20. Similarly, \cite{hoffmann2026ihaloconstrainingmilky} find an average  $\mathrm{DM_{MW,CGM}}$ roughly twice as large as the YT20 model, and \cite{liu2026investigatinganisotropydispersionmeasure} find that the average $\mathrm{DM_{MW,CGM}}$ can significantly exceed the YT20 model average for portions of the sky. Thus we take the YT20 model as our conservative estimate. We include a scatter of 0.2\,dex to capture sightline-to-sightline variance. We emphasize that a factor of 2 increase in $\mathrm{DM_{MW,CGM}}$ would lead to a negligble $\mathrm{DM_{host,CGM}}$ in most cases. This would make the constraints on $\mathrm{DM_{host,CGM}}$ more severe, strengthening the conclusions of this work. On the other hand, if $\mathrm{DM_{MW,CGM}}$ 
were $50\%$ of our assumed values, this would add 
just $\sim15$\,pc\,cm$^{-3}$ to inferred host CGM.

\subsection{Looking ahead - All-Sky Monitors}
\label{sec:asm}

Our method for measuring the CGM in extragalactic halos relies on two characteristics of the FRB sample: low redshift, such that the CGM signal is not washed out by $\mathrm{DM_{cosmic}}$, and negligible $\mathrm{DM_{host,ISM}}$. The latter cannot be predicted, and merely requires a large enough sample such that a good number of FRBs originated from favorable environments (outside the disk, for example). However, with a large enough sample, a $\mathrm{DM_{host,ISM}}$ component could be fit simultaneously to the data. For the former, a large sample of local universe FRBs is expected from upcoming all-sky monitors. The mean distance of an observed sample of FRBs scales as the $1/\sqrt{S_{min}}$, where $S_{min}$ is the minimum detectable flux of the survey \citep{dongzi2019}. All-sky, modest sensitivity telescopes will preferentially discover bright, nearby FRBs with small DM contributions from the IGM. A large FOV is easily achieved at radio frequencies with aperture arrays. This fact will be taken advantage of by the Coherent All-Sky Monitor (CASM) \citep{connor2026256antennacoherentallskymonitor} and similar projects (BURSTT; \cite{Lin2022} and CHARTs in Chile) to discover a large number of local universe FRBs. CASM is expected find order 50 local universe FRBs each year. A sample of this size could be used to build upon the work presented in this paper, potentially discriminating between feedback scenarios in simulations and pinning down the trend in $f_\mathrm{gas}$ across halo masses. As demonstrated in Section \ref{sec:mcmc}, a larger sample will enable constraints not only on $f_\mathrm{gas}$ but also the radial profile of the gas. We also point out that existing and upcoming interferometers like CHIME, ASKAP, DSA-110, CHORD \citep{Vanderlinde}, and the full DSA \citep{Hallinan} will continue to grow the sample of FRBs with limited IGM contribution. While less wide-field than aperture arrays like BURSTT and CASM, these telescopes will find low-luminosity, nearby FRBs by virtue of their high total detection rate and  point-source sensitivity.

Further, a large sample of local universe FRBs can be used to characterize the CGM of the MW as well as external galaxies. Beyond simply the total integrated DM of the MW halo, it will be interesting and feasible to look for deviations from spherical symmetry. The largest will be due to the large and small Magallenic clouds, and the M31 halo (if this is not counted separately from the MW halo). We can likely expect each of these to contribute tens of pc cm$^{-3}$ to the DM in their respective directions \citep{ProchaskaZheng2019}. Thus a first target will be a dipole asymmetry between the northern and southern galactic hemispheres. The recent work by \cite{liu2026investigatinganisotropydispersionmeasure} claims to have detected a dipole anisotropy in the MW CGM in galactic longitude, but more work is required to verify. Furthermore, the X-ray data on the MW suggest that there is a significant disk-like component to the CGM, distinct from the hot volume-filling phase \citep{Yamasaki_2020}. This component likely makes up a significant fraction of the total CGM DM, but separating out its directional dependence will require a large sample. Finally, recent work indicates that the stellar and dark matter components of the MW halo are ellipsoidal and tilted with respect to the stellar disk \citep{Han_2022,Han_2022b}. It would be interesting to determine what impact, if any, this has on the gas distribution in the halo. Several hundred local universe FRB sightlines from CASM is an important prerequisite for this science.

\section{Conclusion}

In this work we have used FRBs to study the CGM of their host galaxies. We use local universe FRBs ($z<0.2$), so that the contribution to the DM from the cosmic web does not dominate the contribution from the host galaxy. We select a sample of five FRBs based on their scattering timescales, rotation measures, host galaxy inclinations, and impact parameters. These FRBs likely have a negligible DM contribution from the host ISM and circumburst medium. We isolate the DM of the host CGM by robustly estimating the DM of the MW and the cosmic web along each sightline. We compare these values to simulations and convert them to measurements of the gas content of the host halos. Our main findings are:

\begin{itemize}
    \item FRBs can measure the gas in halos of mass $10^{11-13}\,M_\odot$, and up to $\sim10^{14}\,M_\odot$ with a larger sample including cluster hosts. In contrast, other observables of diffuse gas are limited in the mass ranges they can probe. Constraining the gas content in a wide range of halo masses is important for understanding baryonic feedback.
    \item The massive elliptical host galaxy of FRB 20240209A is extremely gas poor ($M_\mathrm{gas,vir}=0.02^{+0.02}_{-0.02}M_\mathrm{h,vir}$). Considering multiwavelength data of the host, we find evidence that the gas was expelled by a period of strong AGN feedback. This measurement effectively rules out a no AGN feedback scenario by being in strong tension with the SIMBA simulation with AGN feedback turned off. It also aligns more closely with studies using other observables (X-ray, kSZ) that find strong AGN feedback scenarios. 
    \item The other four FRBs in our sample are broadly consistent with simulations. They hint at a trend of increasing $f_\mathrm{gas}$ when moving to lower mass halos from $M_\mathrm{h}\sim10^{13}$, which is expected from AGN feedback affecting the highest mass galaxies. They are strongly inconsistent with a scenario in which halos retain all of their baryons and trace the dark matter's NFW profile. 
    \item Upcoming aperture array all-sky monitors, such as CASM, will uncover a large number of local universe FRBs suitable for this science. With a larger sample, we will be able to discriminate between AGN feedback implementations in simulations and determine the trend in $f_\mathrm{gas}$ with halo mass to better precision. We will also be able to constrain the radial profile of the gas and the gas in the MW halo. This opens a promising new path to understanding feedback with FRBs.
\end{itemize}

\begin{acknowledgments}

SM thanks Will Golay for a helpful discussion about radio spectra. 

\end{acknowledgments}

\software{astropy \citep{2013A&A...558A..33A,2018AJ....156..123A,2022ApJ...935..167A}, COLOSSUS \citep{Colossus}, emcee \citep{Foreman_Mackey_2013}, yt \citep{yt}, corner \citep{corner}}

\bibliography{sample701}{}
\bibliographystyle{aasjournalv7}

\end{document}